# Discrete FRFT-Based Frame and Frequency Synchronization for Coherent Optical Systems

Oluyemi Omomukuyo, Shu Zhang, Octavia Dobre, Ramachandran Venkatesan, and Telex M. N. Ngatched

*Abstract*—A joint frame and carrier frequency synchronization algorithm for coherent optical systems, based on the digital computation of the fractional Fourier transform (FRFT), is proposed. The algorithm utilizes the characteristics of energy centralization of chirp signals in the FRFT domain, together with the time and phase shift properties of the FRFT. Chirp signals are used to construct a training sequence (TS), and fractional cross-correlation is employed to define a detection metric for the TS, from which a set of equations can be obtained. Estimates of both the timing offset and carrier frequency offset (CFO) are obtained by solving these equations. This TS is later employed in a phase-dependent decision-directed least-mean square algorithm for adaptive equalization. Simulation results of a 32-Gbaud coherent polarization division multiplexed Nyquist system show that the proposed scheme has a wide CFO estimation range and accurate synchronization performance even in poor optical signal-to-noise ratio conditions.

## I. Introduction

COHERENT optical technology has been actively investigated in recent years as a promising technique for next-generation high-capacity transport networks. Current state-of-the-art coherent optical systems utilize digital signal processing (DSP) to compensate for various linear impairments in optical transmission such as chromatic dispersion (CD) and polarization-mode dispersion (PMD). In addition, these systems support the use of a combination of multi-level modulation and polarization division multiplexing (PDM) to increase the number of transported bits.

In digital coherent receivers, a static filter is usually employed for bulk CD compensation, while a set of adaptive finite-impulse-response (FIR) filters are used in performing polarization demultiplexing as well as to compensate for time-varying channel impairments such as PMD and the state of polarization [1]. Blind tap adaptation algorithms like the constant-modulus algorithm (CMA) and the multi-modulus algorithm (MMA) are commonly used to update the tap coefficients of the adaptive FIR filters. However, both CMA and MMA are disadvantaged by long convergence time and the singularity problem [2]. To avoid these problems, a training sequence (TS)-based phase-dependent decision-directed least-mean square (DD-LMS) algorithm has been proposed [3]. The DD-LMS algorithm requires accurate frame synchronization to identify the TS prior to adaptive equalization. In addition, the carrier frequency offset (CFO) has to be estimated and compensated for.

For the frame synchronization, the Schmidl and Cox's algorithm [4] can be adapted for coherent optical single-carrier systems as demonstrated by Zhou in [5]. However, as shown in [6], the Schmidl and Cox's algorithm yields frame synchronization errors under poor optical signal-to-noise ratio (OSNR) conditions. For the frequency synchronization, most of the existing methods in the literature depend on using either the *M*-th power operation [7] or a TS [5] to remove the modulated data phase. Notwithstanding, the *M*-th power operation is disadvantaged by large computational complexity, while the accuracy of the Zhou's TS-based algorithm [5] degrades in poor OSNR conditions. A method which does not depend on removing the modulated data phase has been proposed in [8]. However, this method is not modulation-format transparent, and has a small CFO estimation range.

In this letter, we propose an algorithm which utilizes fractional cross-correlation, together with the time and phase shift properties of the fractional Fourier transform (FRFT), to carry out joint frame and frequency synchronization. Recently, the FRFT has also been proposed for joint synchronization for coherent optical OFDM [9]. However, the method in [9], which utilizes only the FRFT time and phase shift properties, yields frame synchronization errors even in the absence of noise. In addition, this method has a CFO estimation range of ±4 GHz, and it needs the Schmidl and Cox's algorithm to compute the CFO. The proposed scheme is robust to amplified spontaneous emission (ASE) noise, and has a wide CFO estimation range. The proposed technique is demonstrated by means of simulations in a 32-Gbaud 16-ary quadrature amplitude modulation (16-QAM) coherent PDM system.

## II. Operation Principle

The FRFT is a generalization of the conventional Fourier transform through an angle parameter $\phi$ and an order parameter $\alpha$ [10]. For each value of $\phi$, the $\alpha$th-order FRFT rotates a time-domain signal counterclockwise by $\phi$ [11]. In general, we can relate $\phi$ and $\alpha$ as follows [10]:

$$\phi = \frac{\alpha\pi}{2}. \tag{1}$$

The $\alpha$th-order FRFT of a signal $f(t)$ can be defined as [12]:

$$(\mathcal{F}^\phi f)(u) = \int_{-\infty}^{\infty} f(t) K_\phi(t,u) dt, \qquad (2)$$

$$K_\phi(t,u) = \sqrt{\frac{1-j\cot\phi}{2\pi}} exp\left[j\left(\frac{u^2+t^2}{2}\cot\phi - ut\csc\phi\right)\right], \quad (3)$$

where $\mathcal{F}^\phi$ is the FRFT operator associated with angle $\phi$, and $K_\phi(t,u)$ is the transform kernel, defined in (3) for values of $\phi$ that are not multiples of $\pi$. There are several discrete computational algorithms for the FRFT, but in the proposed scheme, we make use of the algorithm in [10] because of its computational efficiency $\left(O(N\log N)\right)$ for an $N$-length signal.

### A. Training Sequence Design

The TS used to perform the joint synchronization is obtained from two discrete-time linear chirp signals with different chirp rates. For simplicity, we consider a finite-duration discrete-time linear chirp with zero initial phase and a center frequency of 0 Hz, which can be expressed as:

$$f[n] = exp[j\pi(2\beta n^2 T^2)] \qquad 0 \leq n \leq N_s - 1, \qquad (4)$$

where $2\beta$ is the chirp rate, $T$ is the sampling period, and $N_s$ is the number of discrete samples. The optimum angle, $\phi_{opt}$, at which the FRFT of the chirp signal yields an impulse is [13]:

$$\phi_{opt} = -\tan^{-1}\left(\frac{1}{2\beta N_s T^2}\right). \qquad (5)$$

In designing the TS, two different values of $\phi_{opt}$ are selected, and the corresponding values of the chirp rates are obtained from (5). These chirp rates are then used in (4) to construct the actual chirp signals. Since the constellation points of the chirp signals lie in a unit circle, the chirp signals are "sliced" and converted into 4-QAM symbols using the method in [14].

At the receiver, the chirp signals are detected by performing the fractional cross-correlation of the received TS and the transmitted one. This operation yields two impulses whose peaks would shift by different amounts depending on the values of the frame offset and CFO.

### B. Joint Frame and Frequency Synchronization

For each polarization, the received symbols are divided into $B$ blocks, each of length $N_s$. To detect chirp signal 1, for each block, we define a detection metric $R_{1b}(u)$, obtained from the fractional cross-correlation [11] of the block with the original transmitted chirp signal 1 as follows:

$$R_{1b}(u) = \left|\mathcal{F}^{-\frac{\pi}{2}}\left[(\mathcal{F}^{\phi_1'} P_{1b})(u)(\mathcal{F}^{\phi_1'} A)^*(u)\right]\right|^2, \qquad (6)$$

where $\phi_1' = \phi_{1opt} + \frac{\pi}{2}$, and $\phi_{1opt}$ is the optimal angle for chirp signal 1, $P_{1b}(u) = r_x(bN_s + u)$, $r_x$ represents the discrete received time-domain samples, $b = 0,1,\cdots,B-1$ is the block index, $u = 0,1,\cdots,N_s - 1$, $A(u)$ are the discrete samples corresponding to the transmitted chirp signal 1, and $*$ is the complex conjugation operation. For each block $b$, the FRFT sample index, $\hat{u}_{1b}$, where $R_{1b}(u)$ has its peak value is:

$$\hat{u}_{1b} = \arg[max_u(R_{1b}(u))]. \qquad (7)$$

We select the specific block $\hat{b}$ at which the maximum value of the detection metric is obtained using the following rule:

$$\hat{b} = \arg[max_b(R_{1b}(\hat{u}_{1b}))]. \qquad (8)$$

The peak shift for chirp signal 1, $\Delta_{n1}$, is then obtained as:

$$\Delta_{n1} = \hat{u}_{1\hat{b}} - \frac{N_s}{2}, \qquad (9)$$

where $\hat{u}_{1\hat{b}}$ is the value of $\hat{u}_{1b}$ corresponding to block $\hat{b}$. The peak shift for chirp signal 2, $\Delta_{n2}$, is obtained in a similar manner. A time shift $\Delta t$ and phase shift $\Delta f$ of a signal in the time domain correspond to shifts of $\Delta t \cos\phi$ and $\Delta f \sin\phi$ in the FRFT domain, respectively [12]. We can then construct the following set of equations which governs the peak shifts:

$$\Delta_{n1} = \Delta t \cos\phi_{1opt} + \Delta f \sin\phi_{1opt},$$
$$\Delta_{n2} = \Delta t \cos\phi_{2opt} + \Delta f \sin\phi_{2opt}. \qquad (10)$$

The solution of (10) is:

$$\Delta t = \frac{\Delta_{n1} \sin\phi_{2opt} - \Delta_{n2} \sin\phi_{1opt}}{\sin(\phi_{2opt} - \phi_{1opt})},$$
$$\Delta f = \frac{\Delta_{n2} \cos\phi_{1opt} - \Delta_{n1} \cos\phi_{2opt}}{\sin(\phi_{2opt} - \phi_{1opt})}. \qquad (11)$$

The frame offset estimate, $\hat{\mu}$, and the CFO estimate, $\hat{\gamma}$, are:

$$\hat{\mu} = \text{round}(\Delta t) + \hat{b} N_s, \qquad (12)$$

$$\hat{\gamma} = \frac{\Delta f}{N_s} R_s, \qquad (13)$$

where round($\cdot$) rounds towards the nearest integer, and $R_s$ is the symbol rate. As shown in (11) and (13), the frequency resolution of the CFO estimation, would depend on $R_s/N_s$, $\phi_{1opt}$, and $\phi_{2opt}$. It can also be deduced from (10)-(13) that the CFO estimation range, $\gamma_{\max}$, of the proposed algorithm is:

$$\gamma_{\max} = \\ \pm\left[\frac{\cos\phi_{1opt}\left(\frac{N_s}{2}-1-\Delta t\cos\phi_{2opt}\right) - \cos\phi_{2opt}\left(\Delta t\cos\phi_{1opt}-\frac{N_s}{2}\right)}{\sin(\phi_{2opt}-\phi_{1opt})}\right]\frac{R_s}{N_s}. \quad (14)$$

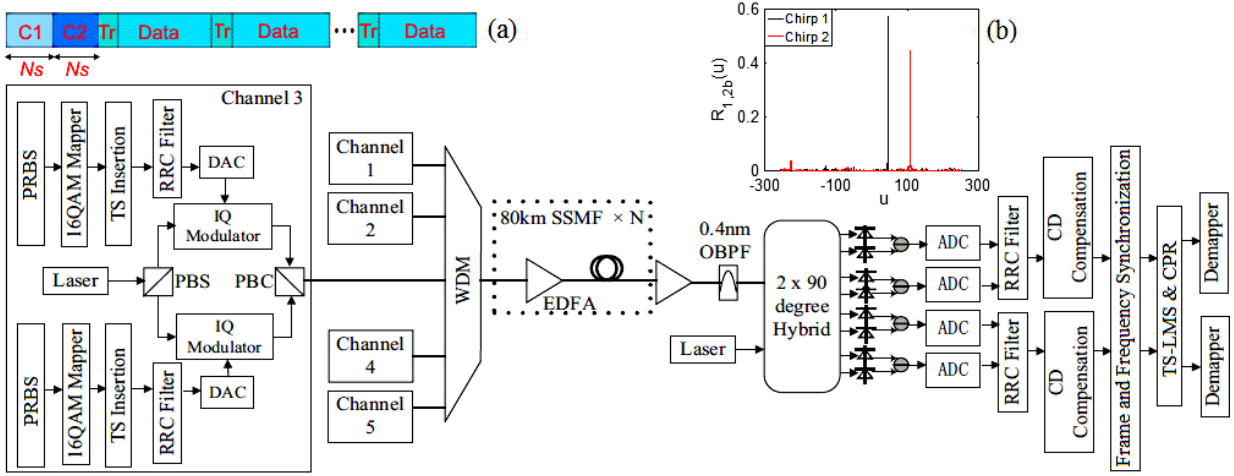

Fig. 1. Simulation setup. C1: chirp signal 1. C2: chirp signal 2. Tr: Training symbols (inserted every 1000 data symbols). PRBS: pseudo-random binary sequence. TS: training sequence. RRC: root raised-cosine. DAC: digital-to-analog converter. IQ: in-phase/quadrature phase. PBS: polarization beam splitter. PBC: polarization beam coupler. WDM: wavelength division multiplexing. EDFA: Erbium-doped fiber amplifier. SSMF: standard single-mode fiber. OBPF: optical band-pass filter. ADC: analog-to-digital converter. CD: chromatic dispersion. LMS: least-mean square. CPR: carrier phase recovery. Inset (a): Frame structure. Inset (b): Metric for both chirp signals using (6) for a 100-symbol frame offset and a 3-GHz CFO.

## III. SIMULATION SETUP AND RESULTS

To investigate the performance of the proposed scheme, a model of a 32-Gbaud coherent PDM Nyquist system, whose schematic is depicted in Fig. 1, is built using VPI TransmissionMaker. Five channels are simulated with a channel spacing of 32 GHz, and the performance is assessed on the central channel. The DSP at the transmitter and receiver is performed in MATLAB. Two independent pseudo-random binary sequences are generated for the two polarization branches. For each polarization, an identical TS, comprising two chirp signals with different chirp rates, is placed at the beginning of each frame to be transmitted to achieve the joint synchronization. An additional 24 training symbols are inserted every 1000 transmitted data symbols to track the dynamic channel behaviors. The frame structure is shown in inset (a) of Fig. 1. The symbols are upsampled to 2 samples/symbol, and digitally shaped using a 73-tap root raised-cosine (RRC) filter with a roll-off factor of 0.13.

For each channel, the electrical signals from each polarization branch are fed to digital-to-analog converters, and then used to drive two null-biased I/Q modulators. The optical source to the I/Q modulators is a continuous wave laser with a linewidth of 100 kHz. The multiplexed PDM optical signal is launched into a transmission link consisting of 10 spans of standard single-mode fiber, with 80 km and a 16-dB gain erbium-doped fiber amplifier per span. At the receiver, the central channel is selected using a 0.4-nm optical band-pass filter, and coherently detected with a polarization-diversity optical hybrid. A laser with a linewidth of 100 kHz is used as the local oscillator. The coherently-detected signal is sampled by the analog-to-digital converters, and then processed by the matched RRC filters. An overlapped frequency-domain equalizer is used for CD compensation. After CD compensation and downsampling to 1 sample/symbol, joint synchronization is carried out using the proposed algorithm.

The processing of the algorithm is carried out independently for each polarization. The TS is then used for polarization demultiplexing using the phase-dependent DD-LMS algorithm [3]. The DD-LMS algorithm is also used to estimate the carrier phase and for residual CFO compensation.

For all simulation results, unless otherwise mentioned, the TS length is 1024, the frame offset is 100 symbols, the CFO is 3 GHz, and the OSNR is 10 dB. In addition, 1000 trial runs have been performed for each assessment. It is clear from (11) and (14) that the performance of the proposed scheme depends on the selection of appropriate values of the angle parameters $\phi_{1opt}$ and $\phi_{2opt}$ in the design of the TS. For the performance assessment, we have selected $\phi_{1opt} = -\phi_{2opt}$.

Inset (b) of Fig. 1 shows that the metric for both chirp signals is impulse-shaped. Fig. 2 shows the frame synchronization performance as a function of $\phi_{2opt}$. It is observed that the timing estimation error is minimum around $\phi_{2opt} = \pi/4$. Consequently, we have carried out the simulations using this value of $\phi_{2opt}$. Fig. 3 shows the impact of a variation of the TS length on the frame and frequency synchronization performance. It is observed that for a TS length of 1024, no timing estimation errors are observed, and the CFO estimation error is ~7 MHz. The frame synchronization performance is more robust than the frequency synchronization performance to further reduction in the TS length. With the above simulation parameters, $\gamma_{\max}$, as obtained using (14), is ~$\pm16.3$ GHz. Fig. 4 shows that the proposed algorithm can comfortably estimate CFOs as high as $\pm5$ GHz, with a maximum CFO estimation error of ~11 MHz obtained. In Fig. 5, the frame and frequency synchronization performance of the proposed algorithm is compared to the TS-based Schmidl-Cox's [4] and the Zhou's [5] algorithms, respectively, in the presence of varying levels of the OSNR. The proposed algorithm demonstrates superior robustness to ASE noise than both algorithms.

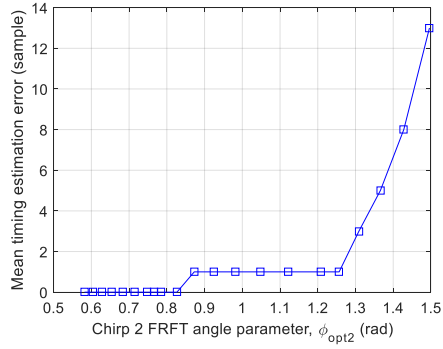

Fig. 2. Frame synchronization performance as a function of $\phi_{2opt}$.

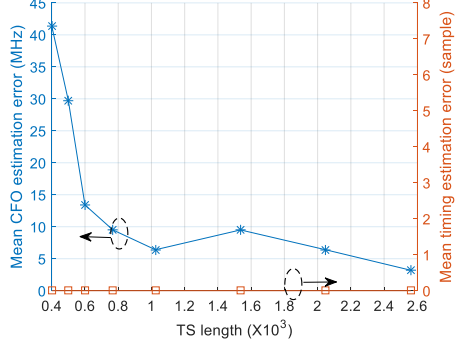

Fig. 3. Frame and frequency synchronization performance as a function of the TS length.

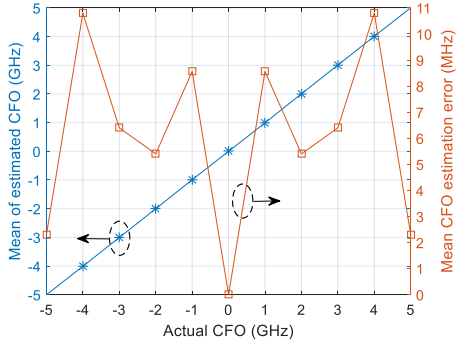

Fig. 4. Mean of estimated CFO and mean of CFO estimation error as a function of the actual CFO.

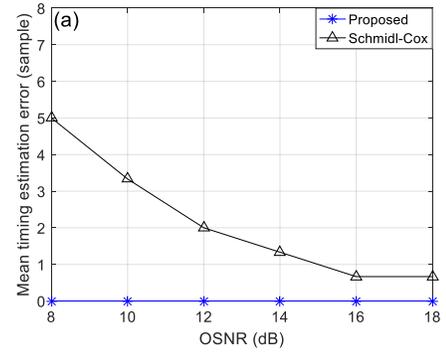

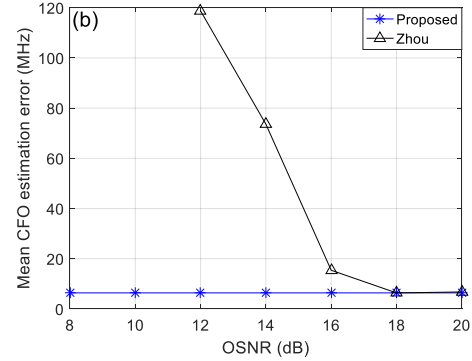

Fig. 5. Synchronization performance in the presence of ASE noise. (a) Frame synchronization. (b) Frequency synchronization.

## IV. CONCLUSION

A novel joint frame and frequency synchronization scheme based on the FRFT has been proposed for coherent optical PDM systems. The proposed scheme, which utilizes fractional cross-correlation, has been shown to be robust to ASE noise, with a wide CFO estimation range, greater than half the symbol rate. The FRFT angle parameter can be varied in the design of the TS in the scheme to increase the accuracy of the offset estimation.